\documentclass{article}

\usepackage[preprint]{neurips_2026}

\usepackage[utf8]{inputenc}
\usepackage[T1]{fontenc}
\usepackage{hyperref}
\usepackage{url}
\usepackage{booktabs}
\usepackage{amsfonts}
\usepackage{amsmath}
\usepackage{amssymb}
\usepackage{nicefrac}
\usepackage{microtype}
\usepackage{xcolor}
\usepackage{graphicx}
\usepackage{float}

\title{What does self-supervision add to engineered features\\for stellar light curves?}

\author{%
  Zijun Liu\thanks{Corresponding authors. Z.L. conceived the project and the encoder, designed and
  ran all experiments, assembled the data and wrote the manuscript. C.A.T. advised on the
  astrophysics, including probe tasks, data sources and label quality. Y.M. advised on the machine-learning methodology, including the
  reconstruction artifact, additional controls and interpretation of the results. C.A.T. and Y.M.
  commented on the manuscript.}\\
  University of California, San Diego\\
  \texttt{brl055@ucsd.edu}
  \And
  Yue Ma\footnotemark[1]\\
  Hal\i c\i o\u{g}lu Data Science Institute\\
  University of California, San Diego\\
  \texttt{y5ma@ucsd.edu}
  \And
  Christopher A. Theissen\footnotemark[1]\\
  Department of Astronomy and Astrophysics\\
  University of California, San Diego\\
  \texttt{ctheissen@ucsd.edu}
}

\begin{document}

\maketitle

\begin{abstract}
TESS light curves encode a star's pulsation, rotation, binarity and oscillations, among other physical
properties, but the survey
produces far more light curves than can be labelled by hand. Astronomers have engineered a few
dozen physically motivated features from them, which together remain a strong baseline. Previous work has trained
encoders on light curves and judged them by their own downstream scores, but has not asked what
they add to the engineered features. We pre-train a convolutional encoder on unlabelled TESS light curves to predict each
window's latent representation from those of its neighbours. We then freeze it, add its features to
the 25 engineered features, and read the combination out with one fixed linear probe on ten tasks:
five variability classes, four asteroseismic targets and rotation period. Adding the learned features
improves the readout on 7 of 10 tasks over the engineered features alone, where an untrained encoder
does not. The combination also outperforms two networks trained end-to-end on the same labelled
stars on most tasks. An encoder trained without labels therefore gives astronomers new features
that complement the ones they already use rather than replace them.
\end{abstract}

\section{Introduction}

A star's light curve carries its pulsation, rotation, eclipses and oscillations, and with them its
age, interior and companions. Wide-field surveys now produce light curves far faster
than they can be labelled. TESS
\citep{ricker2015} returns 2-minute or 20-second \citep{huber2022} light curves for hundreds of
thousands of pre-selected stars, and full-frame images of every star in the field at cadences that
change between mission cycles (30 minutes, then 10, now 200 seconds). LSST \citep{ivezic2019}
will add \emph{billions} more light curves, albeit at a cadence of days. Self-supervised learning is the
natural response: pre-train one encoder on the unlabelled archive, then attach a cheap readout per
task. Such encoders exist for ground-based, Gaia and Kepler light curves
\citep{donoso2025astromer2,zuo2025falco,rui2026jepa} and are scored by classification accuracy,
sometimes against engineered features \citep{rui2026jepa}; a label-free TESS representation
\citep{poznanski2026} is evaluated by whether repeat observations of one star embed close together.
None measures what they add to engineered features.

In this domain, the baseline that a learned encoder must improve on is already strong. Decades of variable-star
research have distilled light curves into compact engineered summaries: periodogram peak structure,
band-limited power, spectral entropy and robust scatter statistics. The benchmark of \citet{li2025staremb} makes the point
sharply: across seven expert-labelled variable-star classes, no time-series foundation model they
tested strictly surpassed the long-established domain feature baseline on classification. A learned
encoder for this data therefore has to answer one question: do its features add predictive signal to
the engineered features under the same readout?

We measure that directly. We pre-train an encoder on TESS light curves, freeze it, and add its
features to the 25 engineered features under one fixed linear readout. Our contributions are:

\begin{itemize}
\item \textbf{The learned features add to the engineered features rather than replacing them.}
Adding the learned features to the engineered ones improves a fixed linear readout beyond
$2\,\mathrm{SE}$ on 7 of 10 tasks, on each of which an untrained encoder gives no such gain, so the gain comes
from pre-training, not from merely adding 128 columns.
\item \textbf{The combined pipeline, read out by a fixed linear model, outperforms the two
supervised networks we train end-to-end on the same labels on most tasks.} Against a supervised Conv1D with the
encoder's architecture it is ahead on 7
of 10 tasks and indistinguishable on the other 3; against an MLP on raw flux it is ahead on 8 of 10.
\end{itemize}

\section{Data and method}

\paragraph{Data.} We use 2-minute SPOC PDCSAP light curves of stars brighter than TESS magnitude
$T = 10$ from Sectors 27--101. Pre-training and three of the ten probes (pulsating, eclipsing
binary and transit) use a label-enriched subset of 13{,}470 stars, split 70/15/15
by star: every star labelled as a transit host, eclipsing binary or pulsator that has a usable
window, plus a random sample of 10{,}000 quiet stars, listed in none of our variability catalogues. The pre-training
loss never sees a label, but this pool is label-enriched, not a random draw from the survey. Five probes draw their labels from catalogues that barely overlap this subset and are scored on a second pool of 22{,}860 stars built the same way. The two rotation probes are scored
on the 106{,}284 corpus stars outside the first pool, split 70/15/15, where rotators are a natural
13\% (Appendix~\ref{app:rotpool}). No test star of the first pool and no star of the rotation pool
enters pre-training. About 8\% of the second pool's test stars (265 of 3{,}429) were in the pre-training subset; removing them leaves every gain in place
(Appendix~\ref{app:unseen}). Each light curve is split into segments wherever the time axis
breaks, at a missing cadence or a jump longer than five median cadences; segments are
median-normalised by their median absolute deviation and sliced into non-overlapping windows of 256
cadences ($256 \times 120$\,s $\approx 8.5$\,h). Windows with a missing value are discarded, never interpolated or padded,
and segments are never stitched across sector or gap boundaries. Each pre-training example is 16
consecutive windows ($\approx 5.69$\,d) starting at a random point in a segment.

\paragraph{Encoder.} A 1D convolutional variational encoder \citep{kingma2014} maps each window to a
128-dimensional Gaussian posterior, and a GRU \citep{cho2014} predicts the next window's latent from
its predecessors, run both forward and backward over the sequence. The objective is
\begin{align*}
\mathcal{L} = {} & \|\hat{x}-x\|^{2} + \beta\,\mathrm{KL}
+ \lambda\big[\|\hat{z}_{t+1}-\bar{z}_{t+1}\|^{2} + \|\hat{z}_{t-1}-\bar{z}_{t-1}\|^{2}\big] \\
& + w\big[\|\log P_{H}(\hat{x})-\log P_{H}(x)\|^{2} + \|\Delta\hat{x}-\Delta x\|^{2}\big],
\end{align*}
where every term is a mean squared error, $\bar{z}$ the stop-gradient target, $P_{H}$ the Hann-tapered power spectrum, $\Delta$ the first difference, and
$(\beta,\lambda,w) = (0.1, 60, 0.3)$, $\beta$ warmed up linearly over ten epochs. The encoder is trained with six seeds and then frozen; no gradient from any label reaches
it.

\paragraph{Representations and readout.} A star's learned features are $\mu$, the mean, over every
window of its first segment (16 for 95\% of stars, 20 for the rest), of the per-window posterior
means. The engineered baseline is a fixed 25-feature vector modelled on the T'DA variability pipeline
\citep{audenaert2021}, computed on the same first segment, 19 from its periodogram and 6 time-domain statistics; Appendix~\ref{app:features} defines each feature. Readouts are fixed in advance and never
tuned on a task, with inputs standardised using training-split statistics alone: class-balanced $L_{2}$ logistic regression
at $C = 1$ for the detection tasks, ridge for the two regressions with its penalty set by leave-one-out
cross-validation. We compare four
arms, each the same readout on a different feature set: \texttt{features}, $\mu$,
\texttt{features}\,$\oplus\,\mu$, and, as a control, the same fusion built on an \emph{untrained}
encoder; the two supervised baselines below are the fifth and sixth arms.
Appendix~\ref{app:readout} repeats the first three under XGBoost.

\paragraph{Supervised baselines.} Two end-to-end baselines see exactly the same input windows, star
sets and splits. \textbf{C1} reuses the encoder's convolutional trunk unchanged, adds a 128-unit linear layer
and a scalar output, and is trained per task with a class-weighted loss on the mean of the star's window logits, computed
inside the network so that the gradient flows through the mean. \textbf{C2} replaces the convolutional trunk with a
dense one and changes nothing else, so C1 versus C2 isolates the convolutional inductive bias. Both
use one regularisation setting, fixed in advance, for all tasks and select on a validation split no
probe has ever touched.

\paragraph{Tasks and metrics.} We score ten tasks with one probe per physical quantity: pulsating
variables \citep{gao2025}, eclipsing binaries \citep{prsa2022}, rotation and rotation period
\citep{boyle2026}, transits \citep{guerrero2021}, flare activity \citep{seli2025,vida2021},
oscillating red giants and their $\nu_{\max}$ \citep{hon2021}, solar-like oscillators
\citep{hatt2023}, and RGB versus helium-burning discrimination \citep{sreenivas2026}. Here
$\nu_{\max}$ is the frequency of maximum oscillation power. Detection tasks are scored by the area
under the precision--recall curve (PR-AUC), whose floor is the positive-class prevalence rather than
$0.5$; prevalence is therefore quoted with every detection score. RGB versus helium-burning
separates two red-giant populations rather than detecting a rare class, so it is scored by ROC-AUC;
the two regressions are scored by $R^{2}$.

\paragraph{Uncertainties.} All error bars are $2\times$ the standard error of the mean over seeds.
For the $\mu$ arms the seeds are independent pre-training runs (six); for C1 and C2 they are
initialisation and shuffling seeds (six); the engineered arm is
deterministic. The $\mu$ and supervised seeds are unrelated, so differences between those arms are
unpaired, with bands combined as $\sqrt{\mathrm{SE}_a^{2}+\mathrm{SE}_b^{2}}$. A difference counts
only beyond its $2\,\mathrm{SE}$ band. These bands
measure re-training spread only; a paired bootstrap over the test stars (1{,}000 resamples, seeds
averaged; Appendix~\ref{app:bootstrap}) checks that no verdict hinges on the test sample.

\section{Results}

\begin{table}[t]
\caption{Ten-task scorecard. Frozen arms: a fixed logistic or ridge readout, or XGBoost on the same
inputs (Appendix~\ref{app:readout}). Bold: best point estimate of the three arms under each frozen
readout, with no significance test (appendix tables bold only differences beyond $2\,\mathrm{SE}$). $n$: test stars;
prev.: positive-class prevalence. The two rotation rows are scored on the rotation pool at the
survey's own rotator fraction; the other detection rows are case-control.}
\label{tab:scorecard}
\centering
\scriptsize
\setlength{\tabcolsep}{3pt}
\begin{tabular}{llrrrrrrrrr}
\toprule
 & & & \multicolumn{3}{c}{frozen linear readout} & \multicolumn{3}{c}{frozen XGBoost readout} & \multicolumn{2}{c}{end-to-end supervised} \\
\cmidrule(lr){4-6}\cmidrule(lr){7-9}\cmidrule(lr){10-11}
task & metric & $n$ (prev.) & features & $\mu$ & features\,$\oplus$\,$\mu$ & features & $\mu$ & features\,$\oplus$\,$\mu$ & Conv1D & MLP \\
\midrule
pulsating & PR-AUC & 2021 (0.107) & 0.789 & 0.806 & \textbf{0.846} & \textbf{0.845} & 0.821 & 0.834 & 0.816 & 0.748 \\
eclipsing binary & PR-AUC & 2021 (0.097) & 0.742 & 0.771 & \textbf{0.775} & 0.798 & 0.789 & \textbf{0.807} & 0.774 & 0.719 \\
rotation & PR-AUC & 15944 (0.130) & 0.386 & 0.373 & \textbf{0.441} & \textbf{0.518} & 0.436 & 0.515 & 0.409 & 0.420 \\
transit & PR-AUC & 2021 (0.060) & 0.190 & 0.144 & \textbf{0.218} & 0.217 & 0.223 & \textbf{0.228} & 0.134 & 0.109 \\
oscillating giant & PR-AUC & 3429 (0.383) & \textbf{0.920} & 0.854 & 0.918 & \textbf{0.924} & 0.899 & 0.923 & 0.875 & 0.883 \\
solar-like osc. & PR-AUC & 3429 (0.122) & 0.320 & 0.336 & \textbf{0.390} & \textbf{0.470} & 0.409 & 0.463 & 0.396 & 0.275 \\
RGB vs. HeB & ROC-AUC & 161 (0.702) & \textbf{0.758} & 0.660 & 0.708 & 0.746 & 0.725 & \textbf{0.763} & 0.530 & 0.762 \\
$\nu_{\max}$ & $R^2$ & 1313 & 0.831 & 0.802 & \textbf{0.867} & 0.916 & 0.831 & \textbf{0.919} & 0.832 & 0.834 \\
rotation period & $R^2$ & 1318 & 0.718 & 0.636 & \textbf{0.752} & \textbf{0.829} & 0.729 & 0.825 & 0.697 & 0.773 \\
flare activity & PR-AUC & 3429 (0.089) & 0.474 & 0.453 & \textbf{0.524} & 0.562 & 0.485 & \textbf{0.565} & 0.533 & 0.444 \\
\bottomrule
\end{tabular}

\end{table}

\begin{figure}[t]
\centering
\includegraphics[width=\linewidth]{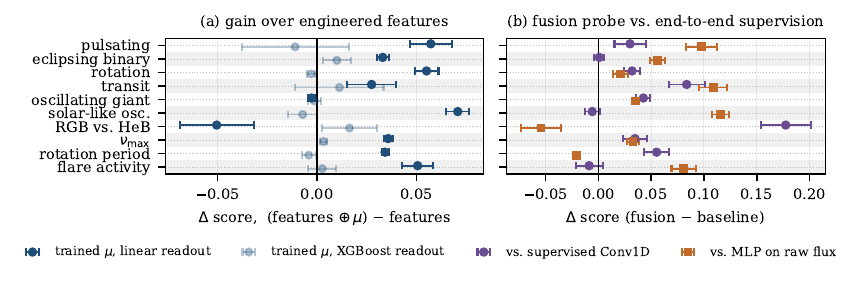}
\caption{The two claims. \textbf{(a)} What the learned features add to the engineered features, under
the fixed linear readout of Table~\ref{tab:scorecard} (solid) and under an XGBoost readout on the
same inputs (faded; Appendix~\ref{app:readout}). \textbf{(b)} The fusion probe against the two
supervised baselines. Bars are $2\,\mathrm{SE}$.}
\label{fig:deltas}
\end{figure}

\paragraph{The learned features add to the engineered features.} Adding $\mu$ to the 25 features
improves the fixed readout (Table~\ref{tab:scorecard}, Figure~\ref{fig:deltas}a) beyond $2\,\mathrm{SE}$ on 7 of 10 tasks, led by solar-like oscillators $+0.071$,
pulsating $+0.057$ and flare activity $+0.051$, and reaching $+0.034$ ($R^{2}$) on rotation period.
Rotation clears the bar too ($+0.055 \pm 0.006$) but does not count: an untrained encoder gains
about half as much there ($+0.027 \pm 0.003$ over six random initialisations), and a gain counts
only if the untrained control's gain stays within $2\,\mathrm{SE}$ of zero, a rule fixed before the
rotation tasks were scored (Appendix~\ref{app:rotpool}). The two negative tasks are those where 128 extra columns cost more
than they add: the 25 features already cover the task (oscillating giants, $-0.003$ on a baseline
of $0.920$) or the training set is too small to fit 128 more weights (RGB versus helium-burning, $-0.050$ on
755 training stars). On both, an untrained encoder loses about as much, within the error band, so
redundant columns, not misleading content in $\mu$, cause the loss (Appendix~\ref{app:negatives}). On the
other nine tasks the untrained control never exceeds $+0.016$ and is negative on six
(Table~\ref{tab:readout}); outside the two rotation tasks it is one random encoder, with no error
band. So the gains that count come from
pre-training, not from 128 extra columns. $\mu$ on its own does not
replace the features (ahead on 3 of 10 alone), as \citet{li2025staremb} find for
foundation models; we therefore report fusion rather than substitution \citep{makinen2024}.

\paragraph{The fusion probe outperforms the two supervised networks on most tasks.} Against C1, which shares
the convolutional trunk, labels and input windows but is trained end-to-end, the fusion
probe is ahead on 7 of 10 tasks and indistinguishable on the other 3
(Figure~\ref{fig:deltas}b). Against
C2, an MLP on raw flux, it is ahead on 8 of 10 and behind on two (RGB versus helium-burning and
rotation period). C1 is not weak: it beats the engineered arm on 5 of 10.
A paired star-bootstrap (Appendix~\ref{app:bootstrap}) excludes zero on 6 of the 7 fusion gains that
count, 6 of the 7 leads over C1 and all 8 over C2.
This holds for this architecture on 755 to 74{,}398 labelled stars per task, not for
supervision in general.

\paragraph{The GRU prediction term raises the gain on every task.} Retraining without it,
all else unchanged, lowers the fusion gain on all 10 tasks, by more than $2\,\mathrm{SE}$ on 6 when
compared seed by seed (Table~\ref{tab:readout}).

\section{Discussion}

\paragraph{Pick the encoder by a held-out probe, not by its loss.} We trained 29 variants of
the pre-training recipe that differ in their loss terms, and scored each on the three detection
tasks of the first star pool, the tasks we developed the recipe on (Appendix~\ref{app:valloss}).
On none of the three did the variant with the lowest validation reconstruction loss score best; choosing it would have cost $0.011$--$0.031$
PR-AUC.

\paragraph{\textbf{Limitations and next steps.}} \emph{Readout.} The four linear arms share one small linear model, which
cannot rescue weak features, so the comparison is between feature sets. Our claim is confined to
these 25 features and a linear readout: under XGBoost the gain exceeds $2\,\mathrm{SE}$ on 3 of 10
tasks and is negative on two, and it shrinks most where XGBoost improves most on the features alone
(Spearman $0.87$; Appendix~\ref{app:readout}). \emph{Data.} Every arm, the supervised baselines
included, sees one 5.7-day segment of a bright 2-minute star; fainter full-frame-image stars and
multi-sector light curves come next. \emph{Labels.} Our
negatives are stars absent from the catalogues, not stars verified to lack the signal, so
catalogue incompleteness adds positive-unlabelled label noise, and detection scores may be
conservative. The flare label marks stars that flared anywhere in their TESS data, so the task is
recognising a flaring star, not finding a flare in the encoded segment. \emph{Label scarcity.} With
the readout trained on a few percent of the labels, the advantage narrows on 7 of 10 tasks, so how
much pre-training helps when labels are scarce remains open. \emph{Reconstruction.} The decoder
plants a narrow spike near the centre of each window; every loss change we tried that removed it
also hurt some probe scores (Appendix~\ref{app:artifact}). \emph{Discovery.} Every task here has a
catalogue to score against; whether $\mu$ also groups stars into kinds of variability no
catalogue lists yet is untested and needs a different evaluation.

\paragraph{Reproducibility and disclosure.} \S2 and Appendices~\ref{app:training}
and~\ref{app:features} give the full recipe; code, configuration files and every score table are at
\url{https://github.com/BrightonLiu-zZ/stellar-ssl/tree/arxiv-v1}. Generative AI assisted with
analysis code and prose editing.

\raggedbottom
\bibliographystyle{plainnat}
\bibliography{refs}

\appendix

\section{Pre-training and readout details}
\label{app:training}

\paragraph{Architecture.} The encoder has four convolutional stages (widths $32/64/128/256$, kernel 5,
each halving length) and a 128-dimensional latent; the predictor is a one-layer GRU of width 256.

\paragraph{What one pre-training example is.} The packed pre-training corpus is the 70\% training
split of the first pool: 9{,}428 stars, 35{,}854 segments and 582{,}352 windows of 256 cadences. The
encoder is monitored on the pool's 15\% validation split (2{,}021 stars, 121{,}328 windows), which no
probe ever sees. One example is one segment. Each epoch it yields a fresh run of 16 consecutive
windows beginning at a uniformly random offset inside that segment, so a long segment shows a
different 5.69-day stretch every epoch; a segment with fewer than 16 windows never entered the pack.
Validation uses offset 0 and every segment, so the monitored quantity does not move with the
sampling. A batch is 256 sequences, giving 140 optimiser steps per epoch.

\paragraph{The first-difference term.} In the objective of \S2, $\Delta x_{t} = x_{t+1} - x_{t}$ is
the first difference along a window, and $\|\Delta\hat{x}-\Delta x\|^{2}$ is the mean squared error
between the reconstruction's cadence-to-cadence changes and the input's. It is there because the
point-wise term $\|\hat{x}-x\|^{2}$ is already small for a smooth curve drawn through a noisy one:
the fastest components of a light curve carry little amplitude, so a decoder that low-passes its
input pays almost nothing for discarding them. Differencing is a high-pass operation, so the same
error re-enters the objective weighted towards high frequency and the smoothed reconstruction is
penalised. $P_{H}$ plays the same role in the frequency domain; the two share the weight $w$ and are
summed with equal weight inside it.

\paragraph{Optimisation.} Adam at learning rate $3\times10^{-4}$, cosine-annealed to
$3\times10^{-6}$ over the 100-epoch cap, gradient-norm clipping at $1.0$, mixed precision. $\beta$
rises linearly from $0$ to its target $0.1$ over the first ten epochs and stays there; the shipped
recipe uses no free-bits floor. Training stops early once ten consecutive epochs improve no tracked
checkpoint, a counter that is armed only after the warm-up.

\paragraph{Which checkpoint is frozen.} Only post-warm-up epochs are eligible. During the warm-up
$\beta$ and the KL term are still moving, and the scheduled-$\beta$ total is minimised at $\beta=0$,
that is by an untrained model, so an unrestricted argmin can select the warm-up transient rather than
a fit. Among the eligible epochs we freeze the one minimising
$\mathrm{recon} + w\cdot\mathrm{aux}$ on the validation split. The prediction term is deliberately
absent from that metric: it is the quantity a dynamics ablation changes, so including it would select
each arm of that comparison by a different rule.

\paragraph{Readouts.} Logistic regression is scikit-learn's \texttt{LogisticRegression} with $L_{2}$
penalty, $C=1$, \texttt{class\_weight="balanced"} and at most 2{,}000 iterations. Ridge is
\texttt{RidgeCV} over ten log-spaced penalties from $10^{-2}$ to $10^{3}$, chosen by leave-one-out
cross-validation. Every input column is z-scored with the training split's own mean and standard
deviation, and the same two numbers are applied unchanged to validation and test. For the fusion arm
the 25 features and the 128 latent means are column-concatenated first and the 153-column matrix is
standardised as a whole, so no arm gets a different scaling of the columns it shares with another.

\paragraph{The supervised baselines.} C1 and C2 are trained with AdamW at learning rate
$3\times10^{-4}$ and weight decay $10^{-4}$, dropout $0.2$ before the task head only (the trunk is
unregularised, as the encoder's is), gradient clipping at $1.0$, mixed precision, batches of 64
stars ($\approx 1{,}040$ windows, since the loss is per star), at most 60 epochs and patience 10.
The positive class enters the loss with weight $n_{\mathrm{neg}}/n_{\mathrm{pos}}$ computed on the
training split. One regularisation setting is fixed in advance for all ten tasks and no per-task
tuning is done. Each run is selected on the best validation score in that task's own headline metric
rather than on validation loss: at the prevalences here the validation-loss minimum and the
validation-metric maximum routinely fall many epochs apart, and selecting on the loss would depress
the baseline we are trying to beat.

\section{The two star pools, and what a quiet star is}
\label{app:pools}

Each task is scored on the pool its catalogue populates, and every arm on a task sees the identical
star list, so the two definitions below fix the population but never a comparison.

\paragraph{The first pool (13{,}470 stars).} Every star in the corpus with at least one usable window
that carries a transit, eclipsing-binary or pulsating label: 3{,}470 stars, assigned to a single
split stratum rarest class first (813 transit, 1{,}226 eclipsing binary, 1{,}431 pulsating). To them
are added 10{,}000 \emph{quiet} stars drawn uniformly at random, without replacement, from the corpus
stars that appear in none of our five v1 catalogues -- transit, eclipsing binary, pulsating, rotation
and flare. The whole pool is split 70/15/15 by star within each stratum, and the encoder is
pre-trained and monitored on this pool's training and validation splits.

Rotation and flare appear here only as exclusions on the quiet draw, so the rotation tasks are scored
on a pool of their own (Appendix~\ref{app:rotpool}).

\paragraph{The second pool (22{,}860 stars).} The four external catalogues behind the asteroseismic
and rotation-period tasks are joined to the corpus into one table of 11{,}007 stars, each row a star
that at least one of them covers. Those 11{,}007, plus the 1{,}853 flare-labelled corpus stars not
already among them, are the pool's members (12{,}860). To them are added a fresh draw of 10{,}000
quiet stars: corpus stars that carry no transit, eclipsing-binary, pulsating or rotation label, have
no flare label, and are absent from the external table. The split is again 70/15/15 by star, with the
stratum taken rarest first over RGB-versus-helium-burning, Kounkel rotation period, flare, solar-like
oscillator, oscillating giant and quiet.

\paragraph{The two screens are not symmetric.} The first pool's quiet draw is screened against the
five v1 catalogues only; it never consults the four external ones, so some of its quiet stars are
oscillating red giants or solar-like oscillators. The second pool's quiet draw is screened against
both sets. Nothing printed here compares a star in one pool with a star in the other, so the
asymmetry costs no verdict; what it does mean is that a negative in either pool is a star
\emph{absent from the relevant catalogues}, not a star examined and found to lack the signal. The
external table is itself a union of four detection lists, so a star it covers with a zero for
oscillating giants is a star Hatt, Sreenivas or Kounkel listed and \citet{hon2021} did not, which is
the same kind of negative. This is the label limitation \S4 states, given exactly.

\section{The two tasks where fusion does not help}
\label{app:negatives}

\begin{center}
\small
\begin{tabular}{lrrrrr}
\toprule
task & features & fusion $\Delta$ & $2\,\mathrm{SE}$ & untrained $\Delta$ & $n_{\mathrm{test}}$ \\
\midrule
oscillating giants & 0.920 & $-0.003$ & 0.002 & $-0.003$ & 3{,}429 \\
RGB vs.\ He-burning & 0.758 & $-0.050$ & 0.019 & $-0.038$ & 161 \\
\bottomrule
\end{tabular}
\end{center}

\paragraph{Oscillating giants: the 25 features already cover the task.} The labels of
\citet{hon2021} come from a convolutional classifier reading a $128 \times 128$ image of each star's
power spectrum, which fires on a visible oscillation power excess at
$6 \lesssim \nu_{\max} \lesssim 250\,\mu$Hz. Our engineered vector encodes close to that same
quantity: eight log-spaced band powers spanning 0.3--360\,c/d, plus the spectral centroid and
entropy, locate the excess and describe its concentration. The features accordingly reach $0.920$,
and the trained and untrained fusion arms then lose the same $0.003$: 128 collinear columns
diluting a linear fit, with no trained-versus-untrained difference left to attribute to $\mu$.

\paragraph{RGB versus helium-burning: no arm resolves the discriminant.} \citet{sreenivas2026} assign
the evolutionary state with a convolutional classifier reading the background-corrected power spectrum
\emph{folded at} $\Delta\nu$; the physics underneath is the dipole mixed-mode period spacing
$\Delta\Pi_{1}$ \citep{bedding2011}. Neither is accessible at our baseline. The encoder sees one
256-cadence window at a time, so the finest frequency structure available to $\mu$ is
$\Delta f = 32.6\,\mu$Hz, while $\Delta\nu$ for these stars is $\approx 1.5$--$18\,\mu$Hz; the comb
itself is unresolved, let alone a fold of it. The engineered features, computed on the whole 5.69\,d
segment, reach $2.0\,\mu$Hz, still coarser than the mixed-mode spacing
$\Delta\Pi_{1}\nu^{2} \approx 0.2$--$0.6\,\mu$Hz near $\nu_{\max} = 50\,\mu$Hz. No arm on this row
measures the quantity the label encodes; the $0.758$ the features reach rides on correlates of
evolutionary state rather than on its discriminant. The row behaves accordingly: it is the only one
where C2 ($0.762$) is ahead of the fusion probe while C1 reaches only $0.530$ on 755 training stars,
and it carries the smallest test set in the study.

\paragraph{Why 128 redundant columns cost a linear fit.} A linear readout spends its training stars
estimating one coefficient per column, and each estimate carries its own error into the prediction
whether or not the column carries signal. On the RGB-versus-helium-burning row the features arm fits
25 coefficients on 755 training stars and the fusion arm 153, so the stars-per-coefficient ratio
falls from about 30 to about 5 while, by the resolution argument above, the added columns supply
nothing the label depends on. The variance a column adds scales with how few stars are available to
pin it down; the signal it adds does not. A column is therefore worth its place only when what it
adds exceeds that cost, and on this row and on oscillating giants it does not. The logistic readout's
$L_{2}$ penalty is fixed at $C = 1$ for every task rather than tuned per task, so it damps this effect
but does not remove it. The prediction that follows is the one the table above shows: an untrained
encoder's 128 columns should cost about the same as a trained one's, since the argument never
mentions what the columns contain, and they do ($-0.038$ against $-0.050$, and $-0.003$ against
$-0.003$).

\section{The gain under a nonlinear readout}
\label{app:readout}

\begin{table}[H]
\caption{What 128 added columns change, relative to the 25 features alone, under each frozen readout
(mean over six encoder seeds; absolute scores are in Table~\ref{tab:scorecard}). $+\mu$: our encoder.
No prediction term: the same recipe trained without the GRU prediction term. Retained: the share of
the full $\mu$ gain that survives that removal, printed only where the full gain is a gain, and
negative where the gain changes sign rather than shrinking. Untrained: a randomly
initialised encoder, which is a single encoder and carries no band. Bold: beyond $2\,\mathrm{SE}$ ---
unlike Table~\ref{tab:scorecard}, where bold marks the best point estimate and implies no test. The
last row counts tasks won, tied and lost at that bar.}
\label{tab:readout}
\centering
\scriptsize
\begin{tabular}{llrrrrrr}
\toprule
 & & \multicolumn{4}{c}{linear readout} & \multicolumn{2}{c}{XGBoost readout} \\
\cmidrule(lr){3-6}\cmidrule(lr){7-8}
task & metric & $+\mu$ & $+\mu$, no prediction term & retained & $+$untrained $\mu$ & $+\mu$ & $+$untrained $\mu$ \\
\midrule
pulsating & PR-AUC & $\boldsymbol{+}\mathbf{0.057}$ & $\boldsymbol{+}\mathbf{0.051}$ & $0.89$ & ${+}0.016$ & ${-}0.011$ & ${-}0.003$ \\
eclipsing binary & PR-AUC & $\boldsymbol{+}\mathbf{0.033}$ & $\boldsymbol{+}\mathbf{0.012}$ & $0.38$ & ${-}0.010$ & $\boldsymbol{+}\mathbf{0.010}$ & ${-}0.003$ \\
rotation & PR-AUC & $\boldsymbol{+}\mathbf{0.055}$ & $\boldsymbol{+}\mathbf{0.036}$ & $0.65$ & ${+}0.029$ & $\boldsymbol{-}\mathbf{0.003}$ & ${-}0.010$ \\
transit & PR-AUC & $\boldsymbol{+}\mathbf{0.027}$ & $\boldsymbol{+}\mathbf{0.018}$ & $0.66$ & ${+}0.004$ & ${+}0.011$ & ${-}0.027$ \\
oscillating giant & PR-AUC & $\boldsymbol{-}\mathbf{0.003}$ & $\boldsymbol{-}\mathbf{0.006}$ & -- & ${-}0.003$ & ${-}0.001$ & ${-}0.008$ \\
solar-like osc. & PR-AUC & $\boldsymbol{+}\mathbf{0.071}$ & $\boldsymbol{+}\mathbf{0.024}$ & $0.33$ & ${-}0.007$ & ${-}0.007$ & ${-}0.026$ \\
RGB vs. HeB & ROC-AUC & $\boldsymbol{-}\mathbf{0.050}$ & $\boldsymbol{-}\mathbf{0.052}$ & -- & ${-}0.038$ & $\boldsymbol{+}\mathbf{0.016}$ & ${-}0.031$ \\
$\nu_{\max}$ & $R^2$ & $\boldsymbol{+}\mathbf{0.036}$ & $\boldsymbol{+}\mathbf{0.010}$ & $0.29$ & ${-}0.001$ & $\boldsymbol{+}\mathbf{0.003}$ & ${-}0.005$ \\
rotation period & $R^2$ & $\boldsymbol{+}\mathbf{0.034}$ & $\boldsymbol{+}\mathbf{0.014}$ & $0.40$ & ${-}0.003$ & $\boldsymbol{-}\mathbf{0.004}$ & ${-}0.013$ \\
flare activity & PR-AUC & $\boldsymbol{+}\mathbf{0.051}$ & $\boldsymbol{+}\mathbf{0.033}$ & $0.65$ & ${+}0.006$ & ${+}0.003$ & ${+}0.006$ \\
\midrule
win / tie / loss & & 8 / 0 / 2 & & & & 3 / 5 / 2 & \\
\bottomrule
\end{tabular}

\end{table}

The retained column is the per-task form of the body's ablation count. On the eight tasks where
fusion gains, removing the GRU prediction term leaves between $0.29$ and $0.89$ of that gain, below
half on four of them; on none does the gain change sign. We give the eight fractions rather
than one summary number because the ten gains are measured in PR-AUC, ROC-AUC and $R^{2}$ and cannot
be added together.

The XGBoost readout takes the same inputs as the linear one. For each task and seed it is selected
over a grid of maximum depth $\{3,5\}$ and learning rate $\{0.05,0.1\}$, with up to 1{,}000 rounds
and early stopping after 50 rounds without improvement on a held-out 25\% of the training stars;
the test stars play no part in the selection. Under XGBoost the features-only arm also varies with
the seed, through the validation split, so the gain is differenced seed by seed.

Under XGBoost the $\mu$ gain (features\,$\oplus\,\mu$ minus features alone) is smaller than under the
linear readout on most tasks. XGBoost also scores higher than the linear readout on the 25 features
alone on nine of ten tasks, and the two changes go together. On solar-like oscillators the features-only score rises from
0.320 to 0.470 and the $\mu$ gain falls from $+0.071$ to $-0.007$. On RGB versus helium-burning,
where XGBoost does not raise the features-only score, the $\mu$ gain is larger under XGBoost. Across
the ten tasks, the rise in the features-only score and the fall in the $\mu$ gain have a Spearman rank
correlation of 0.87. This is consistent with part of the information $\mu$ supplies to a linear
readout being already present in the 25 features, in a form only a nonlinear readout can use.

Table~\ref{tab:readout} also gives the same comparison with an untrained encoder in place of the
trained one. Under XGBoost its 128 columns lower the score on nine of ten tasks and the trained $\mu$
scores above it on eight, so the gains that survive come from pre-training, not from adding columns.
XGBoost on the 25 features alone also scores above the linear fusion probe on 8 of 10 tasks
(Table~\ref{tab:scorecard}). The body's claim is about what $\mu$ adds under a fixed linear readout,
not that the linear fusion probe is the strongest classifier available on these features.

\section{Validation reconstruction against downstream score}
\label{app:valloss}

\begin{figure}[H]
\centering
\includegraphics[width=0.85\linewidth]{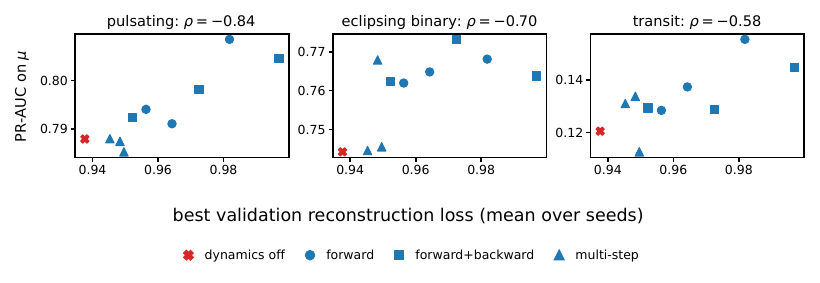}
\caption{Validation reconstruction loss against downstream probe score in the earlier round, one point
per pre-training recipe and one panel per task. The ten recipes share corpus, architecture and loss
terms and differ only in the GRU prediction term (off, forward, forward+backward, multi-step, the last
three at three weights), four seeds each. $x$: best post-warm-up validation reconstruction loss,
averaged over the recipe's seeds; $y$: PR-AUC of the fixed linear probe on $\mu$; $\rho$: Spearman's
rank correlation between negated loss and score. The loss-optimal recipe is the dynamics-free one
(red) on all three tasks, and it is the worst or next-to-worst probe on each.}
\label{fig:valloss_a}
\end{figure}

\begin{figure}[H]
\centering
\includegraphics[width=\linewidth]{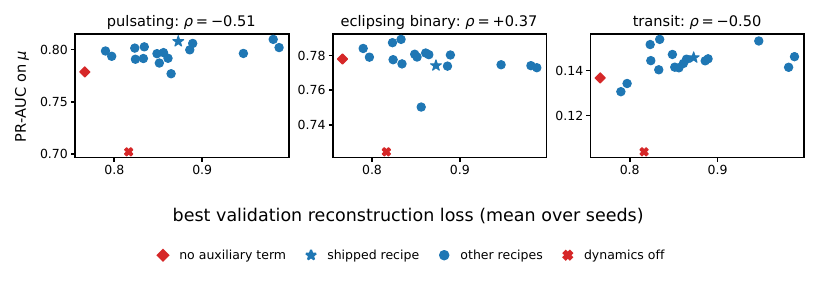}
\caption{The same estimator as Figure~\ref{fig:valloss_a}, on the 19 variants of our final recipe.
The diamond has no auxiliary term, the cross is the one variant without the GRU prediction term, the
star is the recipe we use. $x$: best post-warm-up validation reconstruction loss; $y$: PR-AUC of the
fixed linear probe on $\mu$; $\rho$: Spearman's rank correlation between negated loss and score.}
\label{fig:valloss_b}
\end{figure}

The 29 recipes behind \S4 come from two rounds of pre-training experiments, one per figure.
Figure~\ref{fig:valloss_a} is the earlier round, ten recipes that differ only in the GRU prediction
term; Figure~\ref{fig:valloss_b} is 19 variants of our final recipe. Every recipe was scored on the
three detection tasks of the first star pool, the tasks the recipes were developed on, so those are the
tasks shown. The other seven probes were not run for every recipe, and the rotation tasks are scored
on a pool these recipes were never encoded on. Pooled over all 29 recipes, the one with the lowest
validation reconstruction loss is the same on all three tasks (the diamond in
Figure~\ref{fig:valloss_b}, which has no spectral auxiliary term); it ranks 10th, 27th and 19th of 29
on eclipsing binaries, pulsators and transits, and trusting it costs $0.011$--$0.031$ PR-AUC against
the best recipe. Each figure below repeats the comparison within its own round.

In Figure~\ref{fig:valloss_a} the loss-optimal recipe is the dynamics-free one on all three tasks, and
it is the worst or next-to-worst probe on each, so selecting on the loss costs $0.021$--$0.035$
PR-AUC against the probe-optimal recipe. The trend is not carried by that one point: with it removed,
$\rho$ still runs from $-0.47$ to $-0.87$. Those recipes predate our encoder -- they carry an earlier
spectral auxiliary term -- so Figure~\ref{fig:valloss_b} repeats the estimator on 19 variants of the
final recipe: 16 vary only the form, weight or floor of the spectral auxiliary term, two also put a
floor under the KL term, and one removes the GRU prediction term; six seeds each (two more variants
in which one seed collapsed are excluded; we never drop a seed).

There the anti-correlation does not persist as such: $\rho$ is $-0.51$ on pulsating and $-0.50$ on
transit but $+0.37$ on eclipsing binaries. What persists is that the loss is not
a selector: its optimum is again the recipe with the weakest competing term (the diamond, no
auxiliary term at all), which is never the probe optimum, and trusting it costs $0.011$--$0.031$
PR-AUC. So Figure~\ref{fig:valloss_a} shows an anti-correlation on every task; this one shows it surviving on
two tasks and reversing on one; in neither does the loss choose. Both are statements across recipes
only: within one recipe, across its seeds, no label-free curve metric we tried (reconstruction,
reconstruction plus auxiliary, active latent units) ranks the seeds, every such correlation sitting
within $\pm 0.3$ of zero.

\section{The within-window artifact, and the two gates that governed it}
\label{app:artifact}

\paragraph{Why the objective carries a tapered power spectrum.} A point-wise reconstruction error
prices a light curve by amplitude, and in these data amplitude sits mostly in the slow component, so
a decoder can score well on it while discarding the periodic structure the tasks depend on. The two
auxiliary terms put the pressure back where the point-wise term does not reach: the first difference
in time, and the log power spectrum across frequency. The spectrum is taken on a Hann-tapered window
rather than on the raw one, for the reason the rest of this appendix gives.

\begin{figure}[H]
\centering
\includegraphics[width=\linewidth]{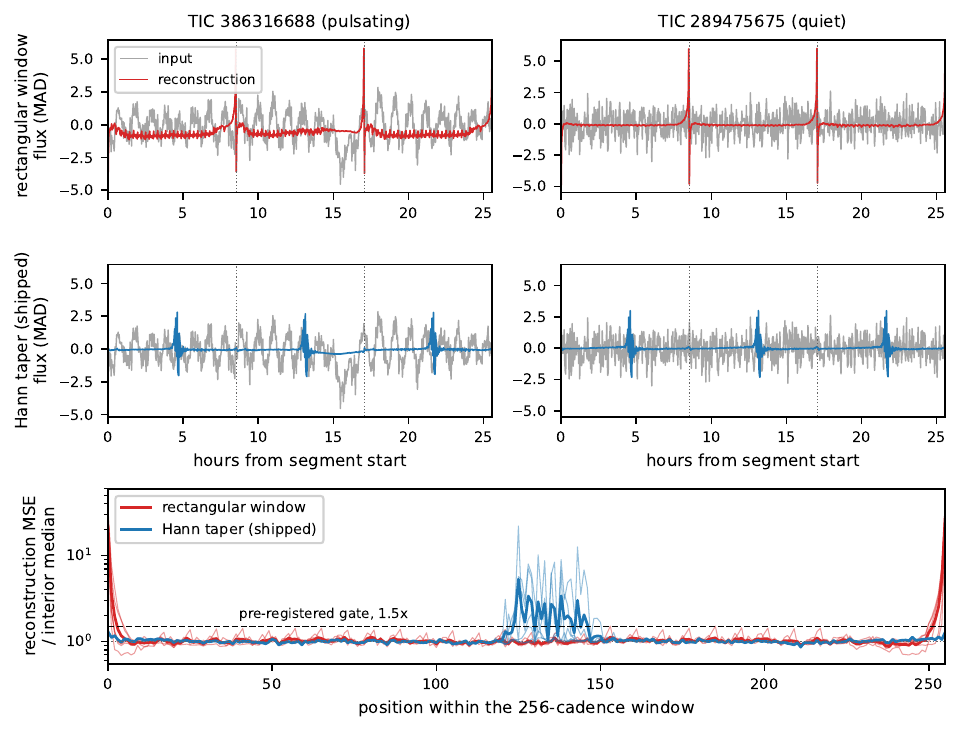}
\caption{The artifact, in two stars and in the average. \textbf{Top two rows:} three consecutive
256-cadence windows of one pulsator and one quiet star, input flux in grey against each recipe's
reconstruction; dotted lines mark the window boundaries, and both rows of a column share one
$y$-scale. The two recipes differ only in the taper used for the auxiliary spectral term. Neither
decoder reproduces the cadence-scale scatter, which a 128-dimensional bottleneck cannot carry; what
differs is where each one puts a spike it does not need. \textbf{Bottom:} reconstruction error by
position within the window, divided by that seed's median over positions 16--240. Thin lines are the
six pre-training seeds, thick lines their mean. The dashed line is the pre-registered artifact gate.
The seed mean understates the tapered recipe's peak because its position moves between seeds
(125--143), which is why the gate is read as a maximum within each seed.}
\label{fig:recon}
\end{figure}

\paragraph{What the decoder found instead.} Every term in the objective is computed inside a single
256-cadence window; nothing spans a boundary. A narrow impulse raises log-power at every frequency at
once, so a decoder can lower the log-spectrum error by planting one spike rather than by reproducing
the star's spectrum. Where the cheapest spike sits is decided by the taper, a fixed function of
position within the window: with a rectangular window it is the window edge, and under a Hann taper
it is the window centre, where the taper weight is largest. Figure~\ref{fig:recon} shows both, in
flux and in the position-resolved error. Laid end to end, the spikes form one train with a period of
exactly one window, which is a property of the display and not of the objective.

Measured as the per-position reconstruction error divided by its own interior median (positions
16--240), maximised over position within each seed and then averaged over six seeds, the rectangular
recipe reaches $31.7\times$ at the window edge and the tapered recipe $11.8\times$ near the centre.
The taper moved the defect and shrank it; it did not remove it. An earlier claim that it had was
based on a measurement taken only at the edge, and is withdrawn.

\paragraph{The two pre-registered acceptance criteria.} Both were registered on 2026-08-15, before
any of the interventions below were written.

\emph{Artifact gate.} The within-window impulse ratio -- the quantity plotted in the lower panel of
Figure~\ref{fig:recon} -- taken as the maximum over \emph{all} positions rather than at the edge
alone, at most $1.5\times$ at full auxiliary weight, over six seeds, \emph{and} a visual sign-off on
a fresh random sample of reconstructed light curves. The number never passes on its own: reading an
edge-only number is how the withdrawn claim above was made in the first place.

\emph{No-regression gate.} No task among eclipsing binaries, pulsators, rotation and transits worse
than the incumbent recipe by more than $2\,\mathrm{SE}$, at both readouts, paired over six seeds. It
is the guard against a clean but useless encoder: a recipe that abolishes the impulse and loses the
probe is not a fix.

\paragraph{What we tried, and why the tapered recipe is the one we ship.} We varied the auxiliary
term along four axes: its \emph{mechanism} (remove it entirely; average the spectrum over a family of
Slepian tapers so that no single position is cheapest; add a penalty on the excess kurtosis of the
residual; both together; put a floor under the spectral sub-term), its \emph{amount} (six weights of
the best mechanism, from $0.0125$ to $0.20$), the \emph{level} of that floor (rebuilt as a two-sided
hinge, which unlike a clamp keeps a restoring gradient, at three measured values), and finally a
\emph{floor on the KL term}, which was the remaining untried lever. No cell cleared both gates. The
ones that removed the impulse lost the probe; the ones that kept the probe kept a measurable impulse.
The nearest miss reached $1.55 \pm 0.05$, which is $2.4\,\mathrm{SE}$ above the gate; its repaired
variant reached $1.47$ over six seeds and $1.55$ over twelve, and regressed on pulsators at both
readouts.

Because no candidate passed, the rule fixed before the ladder ran leaves the incumbent in place. The
encoder this paper uses is therefore the Hann-tapered recipe by a pre-registered fallback, not
because it won a comparison. Two things make that defensible rather than merely procedural. The
artifact was separately measured not to cost the downstream probe -- ablating it moved $\nu_{\max}$
by $+0.0001 \pm 0.0037$ -- and closing the channel, which was also a test of whether the validation
loss becomes a usable selector once it can no longer be lowered by the trick, did not make it one
(Appendix~\ref{app:valloss}).

\section{Paired bootstrap over test stars}
\label{app:bootstrap}

The $2\,\mathrm{SE}$ bands in the body come from re-training and say nothing about whether the
particular test stars favoured one arm. For every printed contrast we therefore resample each task's
test set with replacement 1{,}000 times, rescore every arm on the identical resampled list from its
saved per-star predictions (no model is retrained), average each arm over its seeds, and take the
difference. Table~\ref{tab:bootstrap} gives the 95\% percentile interval of that difference and the
share of resamples in which the fusion arm is ahead; intervals that exclude zero are bold. The check
tightens no verdict and reverses none: every row the body calls ahead has the fusion arm ahead in at
least 85\% of resamples. It does loosen five of the 40 verdicts, each a resolved $2\,\mathrm{SE}$
verdict whose interval touches zero: the gain on transit and the loss on oscillating giants over the
features, the lead over C1 on
pulsating, the loss to C2 on RGB versus helium-burning (161 test stars), and the dynamics contrast on
flare activity.

\begin{table}[H]
\caption{Paired star-bootstrap of the four contrasts the body counts. Each entry is the 95\% interval
of (left arm $-$ right arm) over 1{,}000 resamples of the test stars, seeds averaged, and the share of
resamples with the difference above zero ($f_{>0}$; 1.00 means the left arm is ahead in every
resample, 0.00 that it is behind in every resample). Bold: interval excludes zero.}
\label{tab:bootstrap}
\centering
\scriptsize
\setlength{\tabcolsep}{2pt}
\begin{tabular}{lrrrrrrrr}
\toprule
task & \multicolumn{2}{c}{fusion $-$ features} & \multicolumn{2}{c}{fusion $-$ C1} & \multicolumn{2}{c}{fusion $-$ C2} & \multicolumn{2}{c}{fusion $-$ dynamics-off fusion} \\
 & 95\% CI & $f_{>0}$ & 95\% CI & $f_{>0}$ & 95\% CI & $f_{>0}$ & 95\% CI & $f_{>0}$ \\
\midrule
pulsating & $\mathbf{[+0.012, +0.105]}$ & 0.99 & $[-0.008, +0.073]$ & 0.94 & $\mathbf{[+0.045, +0.150]}$ & 1.00 & $[-0.023, +0.036]$ & 0.69 \\
eclipsing binary & $\mathbf{[+0.005, +0.062]}$ & 0.99 & $[-0.024, +0.025]$ & 0.52 & $\mathbf{[+0.022, +0.094]}$ & 1.00 & $\mathbf{[+0.004, +0.038]}$ & 0.99 \\
rotation & $\mathbf{[+0.043, +0.066]}$ & 1.00 & $\mathbf{[+0.017, +0.048]}$ & 1.00 & $\mathbf{[+0.005, +0.039]}$ & 0.99 & $\mathbf{[+0.014, +0.025]}$ & 1.00 \\
transit & $[-0.025, +0.065]$ & 0.85 & $\mathbf{[+0.041, +0.124]}$ & 1.00 & $\mathbf{[+0.046, +0.174]}$ & 1.00 & $[-0.016, +0.035]$ & 0.78 \\
oscillating giants & $[-0.010, +0.005]$ & 0.21 & $\mathbf{[+0.028, +0.057]}$ & 1.00 & $\mathbf{[+0.021, +0.051]}$ & 1.00 & $[-0.002, +0.008]$ & 0.87 \\
solar-like osc. & $\mathbf{[+0.044, +0.097]}$ & 1.00 & $[-0.037, +0.022]$ & 0.34 & $\mathbf{[+0.079, +0.151]}$ & 1.00 & $\mathbf{[+0.025, +0.067]}$ & 1.00 \\
RGB vs.\ HeB & $\mathbf{[-0.087, -0.012]}$ & 0.00 & $\mathbf{[+0.099, +0.247]}$ & 1.00 & $[-0.126, +0.020]$ & 0.08 & $[-0.037, +0.039]$ & 0.52 \\
$\nu_{\max}$ & $\mathbf{[+0.028, +0.043]}$ & 1.00 & $\mathbf{[+0.024, +0.045]}$ & 1.00 & $\mathbf{[+0.018, +0.047]}$ & 1.00 & $\mathbf{[+0.020, +0.031]}$ & 1.00 \\
rotation period & $\mathbf{[+0.023, +0.046]}$ & 1.00 & $\mathbf{[+0.039, +0.073]}$ & 1.00 & $\mathbf{[-0.038, -0.004]}$ & 0.01 & $\mathbf{[+0.014, +0.027]}$ & 1.00 \\
flare & $\mathbf{[+0.019, +0.078]}$ & 1.00 & $[-0.047, +0.029]$ & 0.32 & $\mathbf{[+0.027, +0.128]}$ & 1.00 & $[-0.000, +0.035]$ & 0.97 \\
\bottomrule
\end{tabular}

\end{table}

\section{The 25 engineered features}
\label{app:features}

All 25 features are computed from a star's first segment: its windows concatenated into one flux
series $x$ of 4{,}096 or 5{,}120 cadences, already median-normalised by its median absolute
deviation. The periodogram is the power spectrum of the mean-subtracted series,
$P(f) = |\mathrm{FFT}(x - \bar{x})|^{2}$ with the zero-frequency bin dropped, and
$p(f) = P(f)/\sum_f P(f)$ its normalisation to unit sum. Frequencies are in cycles per day.
$f_{1}, f_{2}, f_{3}$ are the three strongest local maxima of $P$, strongest first.

\begin{table}[H]
\caption{The engineered feature vector, modelled on the T'DA variability pipeline \citep{audenaert2021}.}
\label{tab:features}
\centering
\small
\begin{tabular}{llp{0.56\linewidth}}
\toprule
\# & feature & definition \\
\midrule
1--3 & $\log_{10} f_{1}, \log_{10} f_{2}, \log_{10} f_{3}$ & frequencies of the three strongest periodogram peaks \\
4--6 & $p_{1}, p_{2}, p_{3}$ & normalised power at each peak, $p_{k} = P(f_{k})/\sum_f P(f)$ \\
7--8 & $p_{2}/p_{1}$, $p_{3}/p_{1}$ & peak-power ratios \\
9 & harmonic ratio & $P(2f_{1})/P(f_{1})$ at the bin nearest $2f_{1}$; $0$ if $2f_{1}$ exceeds the Nyquist frequency \\
10--17 & band powers & $\log_{10} \sum_{f \in B_{j}} p(f)$ for eight log-spaced bands $B_{j}$ spanning 0.3--360\,c/d \\
18 & spectral entropy & $-\sum_f p(f)\ln p(f) / \ln N$, $N$ the number of frequency bins \\
19 & spectral centroid & $\log_{10} \sum_f f\,p(f)$ \\
20--21 & skew, kurtosis & sample skewness and excess kurtosis of $x$ \\
22 & point-to-point scatter ratio & $\mathrm{std}(\Delta x)/\mathrm{std}(x)$, $\Delta$ the first difference \\
23 & depth & 95th minus 5th percentile of $x$ \\
24 & MAD & median absolute deviation of $x$ \\
25 & IQR & inter-quartile range of $x$ \\
\bottomrule
\end{tabular}
\end{table}

\paragraph{What ``modelled on T'DA'' means.} \citet{audenaert2021} classify Kepler Q9 light curves
into eight variability classes with an ensemble of four supervised learners and a random-forest
metaclassifier. We take none of that: not the classifiers, not the classes, not the training set.
What we borrow is the shape of the feature vector its tree-based members read, and the borrowing is
partial in ways a reader should be able to check.

\emph{Kept.} A short list of periodogram peaks with their relative strengths, a harmonic
measure of the dominant peak, an entropy measure of the signal, and robust time-domain scatter
statistics -- skewness, kurtosis and the median absolute deviation, all three of which appear in
their Table~1. Our point-to-point scatter ratio $\mathrm{std}(\Delta x)/\mathrm{std}(x)$ is, up to a
square root, their variability index $\eta_{e}$, the ratio of the mean squared successive difference
to the variance.

\emph{Changed.} They extract up to six frequencies with up to ten harmonics each by iterative
prewhitening -- fit a frequency and its harmonics, subtract them, repeat until a significance
criterion stops the loop. We take the three strongest local maxima of one segment's periodogram, with
no prewhitening and no significance test, and keep one harmonic ratio $P(2f_{1})/P(f_{1})$ instead of
a harmonic series. Their amplitudes and phases become, here, powers normalised by the total and
ratios between them: the flux is median-absolute-deviation normalised per segment, so an absolute
amplitude is not a quantity we have, and we drop phases and phase differences entirely. Their
FliPer, the mean power above each of four frequency cut-offs, becomes our eight log-spaced band
powers spanning 0.3--360\,c/d plus a spectral centroid. Their entropy features are computed on the
time series (differential entropy, multiscale sample entropy); ours is the Shannon entropy of the
normalised power spectrum. Because a packed segment is uniformly sampled and gap-free, its
Lomb--Scargle periodogram is the FFT periodogram, so we compute the cheaper one.

\emph{Dropped.} Everything that needs a phase fold at a reliably determined period -- the
self-organising-map location, the folded point-to-point statistics and the folded range -- because
5.69 days covers few cycles for the longer-period classes in our menagerie. Also dropped: the
coherency parameter, zero crossings, the range of the cumulative sum, the Shapiro--Wilk statistic,
the variance ratio, the significant-harmonic count and the flux ratio, and the power-spectrum image
their deep-learning member reads, which is not a feature vector at all.

The result is 25 numbers, 19 from the periodogram and 6 from the time domain, and it is a baseline
rather than a reproduction: it is deliberately the cheap, linear-readable part of a pipeline whose
published accuracy came from an ensemble we do not run.

\section{Evaluation restricted to stars never seen in pre-training}
\label{app:unseen}

The encoder is fitted on the training stars of the first pool and monitored on its validation stars,
so the test stars of
the three tasks scored on that pool are unseen by construction, and the rotation pool excludes every
pre-training star. The second pool shares some stars with
it. Table~\ref{tab:unseen} removes every test star that was in the pre-training training or validation
split and rescores the saved per-star predictions; no model is refitted. The gain exceeds
$2\,\mathrm{SE}$ on the same tasks as before; the largest change is flare activity, $+0.051$ to
$+0.041$, and the small loss on oscillating giants ($-0.003$) becomes a tie ($-0.001$).

\begin{table}[H]
\caption{The fusion gain (features\,$\oplus\,\mu$ minus features, linear readout) on all test stars and
on the test stars never seen in pre-training. Bold: beyond $2\,\mathrm{SE}$ over six encoder seeds.}
\label{tab:unseen}
\centering
\scriptsize
\begin{tabular}{llrrrr}
\toprule
task & metric & test stars & unseen test stars & gain, all & gain, unseen \\
\midrule
pulsating & PR-AUC & 2021 & 2021 & $\boldsymbol{+}\mathbf{0.057}$ & $\boldsymbol{+}\mathbf{0.057}$ \\
eclipsing binary & PR-AUC & 2021 & 2021 & $\boldsymbol{+}\mathbf{0.033}$ & $\boldsymbol{+}\mathbf{0.033}$ \\
rotation & PR-AUC & 15944 & 15944 & $\boldsymbol{+}\mathbf{0.055}$ & $\boldsymbol{+}\mathbf{0.055}$ \\
transit & PR-AUC & 2021 & 2021 & $\boldsymbol{+}\mathbf{0.027}$ & $\boldsymbol{+}\mathbf{0.027}$ \\
oscillating giant & PR-AUC & 3429 & 3164 & $\boldsymbol{-}\mathbf{0.003}$ & ${-}0.001$ \\
solar-like osc. & PR-AUC & 3429 & 3164 & $\boldsymbol{+}\mathbf{0.071}$ & $\boldsymbol{+}\mathbf{0.069}$ \\
RGB vs. HeB & ROC-AUC & 161 & 149 & $\boldsymbol{-}\mathbf{0.050}$ & $\boldsymbol{-}\mathbf{0.048}$ \\
$\nu_{\max}$ & $R^2$ & 1313 & 1218 & $\boldsymbol{+}\mathbf{0.036}$ & $\boldsymbol{+}\mathbf{0.034}$ \\
rotation period & $R^2$ & 1318 & 1318 & $\boldsymbol{+}\mathbf{0.034}$ & $\boldsymbol{+}\mathbf{0.034}$ \\
flare activity & PR-AUC & 3429 & 3164 & $\boldsymbol{+}\mathbf{0.051}$ & $\boldsymbol{+}\mathbf{0.041}$ \\
\bottomrule
\end{tabular}

\end{table}

\section{The rotation pool}
\label{app:rotpool}

\paragraph{Construction.} Every corpus star with at least one usable window that is not in the
first pool: 106{,}284 stars, of which 13{,}789 carry a rotation period in the TESS All-Sky Rotation
Survey \citep{boyle2026} and 92{,}495 do not. By construction the pool contains no transit host,
eclipsing binary or pulsator (the first pool took every one), no star that entered pre-training, and
no sampling step: rotators are 13.0\% of it, against 12.5\% of the corpus, so both rotation tasks are
scored at the survey's own prevalence rather than the case-control prevalence of the other pools. It
is split 70/15/15 by star, stratified on rotators with period at most five days (8{,}786), other
rotators (5{,}003) and non-rotators, seed 0. The rotation-period task keeps the rotators with period
at most five days. A non-rotator here is a star the survey does not list, not a star examined and
found not to rotate. No encoder is run again: the per-star encoder outputs and the 25 engineered
features for these stars had been cached for an earlier survey-prevalence rescore, and the readouts
are the body's, fitted on this pool's training split. The design, the arms, the readouts and the rule
for reading the result were fixed in writing before any score on this pool was computed.

Two properties still differ between the classes besides rotation, both astrophysical rather than
sampling artefacts: 10.4\% of the rotators and 0.5\% of the non-rotators carry a flare label, and
0.3\% of the rotators against 9.4\% of the non-rotators are oscillating giants or solar-like
oscillators. The headline rows keep every star, because a survey contains them; two sensitivity rows
refit the readout with flare stars removed from both classes, and with the positives restricted to
stars in no non-rotation catalogue and the negatives to stars in no catalogue at all.

\paragraph{The counting rule.} A gain on these two tasks counts only if (i) it exceeds
$2\,\mathrm{SE}$, (ii) the untrained control's own gain stays within $2\,\mathrm{SE}$ of zero, and
(iii) the trained-minus-untrained difference, paired seed $s$ with initialisation $s$, exceeds
$2\,\mathrm{SE}$. Clause (ii) needs the untrained encoder's spread, so six random initialisations
were scored here; Table~\ref{tab:scorecard} prints the first of them, the same initialisation as
every other task's untrained arm.

\paragraph{Result.} Table~\ref{tab:rotpool} gives both tasks with the two sensitivity rows. On
rotation the fusion gain is $+0.055 \pm 0.006$ and the trained encoder is ahead of the untrained one
by $+0.029 \pm 0.008$, but the untrained encoder itself gains $+0.027 \pm 0.003$ over its six
initialisations: with 74{,}398 training stars, 128 random columns are no longer redundant to a linear
fit. Clause (ii) fails, so rotation does not count. On rotation period the gain is $+0.034 \pm 0.002$
on 1{,}318 test stars, the untrained control is at $-0.003$ and the paired difference is
$+0.037 \pm 0.003$, so it counts. The two sensitivity rows agree with the headline on both verdicts.
The dynamics-off contrast is positive on both tasks ($+0.019 \pm 0.007$ and $+0.020 \pm 0.003$), and
a paired star-bootstrap of every contrast in the table (1{,}000 resamples, as in
Appendix~\ref{app:bootstrap}) agrees with every seed-band verdict.

\begin{table}[H]
\caption{The two rotation tasks on the rotation pool. The ``all stars'' rows are those of
Table~\ref{tab:scorecard}; linear and XGBoost readouts and the two supervised baselines are as there.
Gain = features\,$\oplus\,\mu$ minus features, mean over six encoder seeds, bold beyond
$2\,\mathrm{SE}$; untrained gain = the same with the untrained encoder. Sensitivity rows refit the
linear readout on the restricted population. $n$: test stars; prev.: rotator prevalence.}
\label{tab:rotpool}
\centering
\scriptsize
\setlength{\tabcolsep}{2.5pt}
\begin{tabular}{llrrrrrrrrrr}
\toprule
 & & & \multicolumn{5}{c}{linear readout} & \multicolumn{2}{c}{XGBoost} & \multicolumn{2}{c}{supervised} \\
\cmidrule(lr){4-8}\cmidrule(lr){9-10}\cmidrule(lr){11-12}
task & population & $n$ (prev.) & features & $\mu$ & fusion & gain & untr.\ gain & features & fusion & Conv1D & MLP \\
\midrule
rotation (PR-AUC) & all stars & 15944 (0.130) & 0.386 & 0.373 & 0.441 & $\mathbf{+0.055}$ & $+0.029$ & 0.518 & 0.515 & 0.409 & 0.420 \\
 & \quad no flare stars & 15670 (0.119) & 0.356 & 0.352 & 0.413 & $\mathbf{+0.057}$ & $+0.029$ & -- & -- & -- & -- \\
 & \quad strict & 14182 (0.129) & 0.367 & 0.372 & 0.427 & $\mathbf{+0.061}$ & $+0.036$ & -- & -- & -- & -- \\
\midrule
rotation period ($R^2$) & all stars & 1318 & 0.718 & 0.636 & 0.752 & $\mathbf{+0.034}$ & $-0.003$ & 0.829 & 0.825 & 0.697 & 0.773 \\
 & \quad no flare stars & 1174 & 0.713 & 0.628 & 0.744 & $\mathbf{+0.031}$ & $-0.006$ & -- & -- & -- & -- \\
 & \quad strict & 1145 & 0.718 & 0.636 & 0.748 & $\mathbf{+0.031}$ & $-0.004$ & -- & -- & -- & -- \\
\bottomrule
\end{tabular}

\end{table}

\end{document}